\documentclass[twocolumn,aps,prb]{revtex4}

\usepackage{amsmath}
\usepackage{amssymb}
\usepackage{graphicx}
\usepackage{epsfig}
\usepackage{bm}
\usepackage{units}
\usepackage{subfigure}
\usepackage{hyperref}
\usepackage{breakurl}

\newcommand{\be}{\begin{equation}}
\newcommand{\ee}{\end{equation}}
\newcommand{\bea}{\begin{eqnarray}}
\newcommand{\eea}{\end{eqnarray}}
\newcommand{\ben}{\begin{enumerate}}
\newcommand{\een}{\end{enumerate}}
\newcommand{\bit}{\begin{itemize}}
\newcommand{\eit}{\end{itemize}}

\newcommand{\la}[1]{\label{#1}}
\newcommand{\Eq}[1]{Eq.~(\ref{#1})}

\newcommand{\Sec}[1]{Sec.~\ref{#1}}

\newcommand{\Fig}[1]{Fig.~\ref{#1}}

\newcommand{\vv}{\boldsymbol}					

\newcommand{\bert}{\raise-0.45mm\hbox{\Large$\Box$}}

\begin{document}

\title{Irreversibility in an ideal fluid}
 
\author{Alejandro Jenkins}\email{jenkins@hep.fsu.edu}

\affiliation{High Energy Physics, Florida State University, Tallahassee, FL 32306-4350, USA}
\affiliation{Escuela de F\'isica, Universidad de Costa Rica, 11501-2060 San Jos\'e, Costa Rica}

\date{Jan.\ 2013, last revised Mar.\ 2014; to appear in Am.\ J.\ Phys.\ {\bf 82}} 

\begin{abstract}

When a real fluid is expelled quickly from a tube, it forms a jet separated from the surrounding fluid by a thin, turbulent layer.  On the other hand, when the same fluid is sucked into the tube, it comes in from all directions, forming a sink-like flow.  We show that, even for the ideal flow described by the time-reversible Euler equation, an experimenter who only controls the pressure in a pump attached to the tube would see jets form in one direction exclusively.  The asymmetry between outflow and inflow therefore does not depend on viscous dissipation, but rather on the experimenter's limited control of initial and boundary conditions.  This illustrates, in a rather different context from the usual one of thermal physics, how irreversibility may arise in systems whose microscopic dynamics are fully reversible. \\

{\it Keywords:} jets, irreversibility, Euler equation, Machian propulsion \\

{\it PACS:}
47.10.-g,		
47.60.Kz, 		
01.70.+w		

\end{abstract}

\maketitle


\section{Introduction}
\label{sec:intro}

Irreversibility is a prominent feature of real-world phenomena.  If the video of any but the very simplest process is played backwards, it will soon seem so contrary to experience as to appear comical.  For instance, a glass bottle may fall to the floor and shatter into jagged pieces, but we never see the jagged pieces jump up and reassemble themselves into a glass bottle.  This irreversibility is captured by the second law of thermodynamics, which states that the entropy (defined as proportional to the logarithm of the number of microscopic physical configurations that are macroscopically indistinguishable) tends towards a maximum.  Nontechnically, this is often presented as the statement that the randomness (or disorder) of the world increases with time.

As far as we know, the microscopic laws of physics are time-reversal symmetric (at least to a very good approximation), so that nothing actually {\it forbids} the pieces of the broken bottle from reassembling themselves.\cite{CP}  The reason why we never see that happen is that it would require an exquisite tuning of initial conditions: the thermal motions of the molecules in the floor would have to conspire to push the various pieces with just the right velocities to make them converge into the form of the original bottle. If the air, moisture, etc.\ moved out at the right time from the regions of contact between the pieces, the glass would fuse and a pristine bottle may then land on the table above.

The point is that it is very easy for an experimenter to set up the conditions for a bottle to break, but extraordinarily difficult to set up the conditions for the pieces to reassemble themselves:  there are many ways for a bottle to be broken, but comparatively very few for it to be whole.\cite{subjective}  The irreversibility is therefore a consequence of the experimenter's limited control over initial conditions.  For an accessible introduction to this interpretation of irreversibility and to the problem of the ``arrow of time'' in statistical mechanics, see Ref.~\onlinecite{Feynman-time}.

This problem of the arrow of time is usually posed in the context of {\it thermal} systems, in which mechanical energy can be dissipated into the random motion of the particles in a ``heat bath.''  In this article we will describe an idealized, non-dissipative system in which a kind of irreversibility nonetheless appears for the same qualitative reason as in statistical mechanics: because of the experimentalist's limited control over the initial and boundary conditions.  Our example will be the flow of an ideal fluid into and out of a tube.

\section{Machian propulsion}
\label{sec:machian}

\begin{figure}[b]
\begin{center}
	\includegraphics[width=0.4 \textwidth]{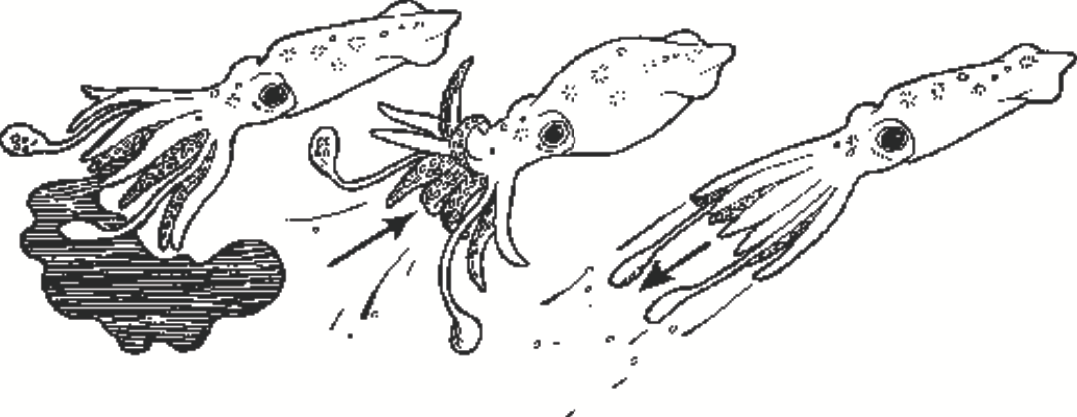}
\end{center}
\caption{A squid escapes from danger by emitting a cloud of ink that obscures it from predators.  It then fills up its mantle cavity with water before rapidly expelling it as a jet.  The cycle of aspiration and expulsion allows the animal to swim away quickly.  The illustration is taken from Ref.~\onlinecite{animals} and is used here with the publisher's permission.\la{fig:squid}}
\end{figure}

In his delightful {\it Physics for Entertainment}, originally published in Russian in 1913, Yakov Perelman expresses surprise that squids, jellyfish, and other aquatic creatures propel themselves by alternately aspirating and then expelling water.  According to Perelman, this method of swimming might appear, at first glance, as absurd as the Baron Munchausen lifting himself out of a swamp by pulling on his own hair.\cite{Perelman}  Even though the water moves in opposite directions during intake and outtake (see \Fig{fig:squid}), a squid can swim in one direction for long periods and at considerable velocity.\cite{SA-squid}

In fact, it is {\it reverse} propulsion during steady aspiration of an ideal fluid that would be akin to the Baron pulling himself up by the hair.  While the outgoing jet efficiently transfers momentum away from the squid and into the surrounding fluid medium, the inflow impinges on the internal cavity and therefore imparts its momentum to the animal.  The only way in which the inflow can end up transferring some of its momentum to the surrounding medium is by viscous diffusion.\cite{Reverse}

Ernst Mach appears to have been the first experimenter to report that if a solid device alternately aspirates and expels fluid through a single opening, it moves in the same direction as if it only expelled the fluid.\cite{Mach}  We shall therefore refer to this process as ``Machian propulsion,'' a term introduced in Ref.~\onlinecite{Machian}.  In France, an equivalent observation has sometimes been dubbed the ``paradox of Bergeron,'' after mechanical engineer Paul Bergeron.\cite{Sedille,Comolet}  The same phenomenon is also notorious because of its role in an often-told story from the early life of the eminent theoretical physicist Richard Feynman.\cite{Feynman-anecdote,Feynman-letter,Wheeler,Creutz}  In certain contexts, engines driven by Machian propulsion have been characterized as ``valveless pulse jet'' devices.\cite{valveless}

\begin{figure}[t]
\centering
\subfigure[]{\includegraphics[width=0.3 \textwidth]{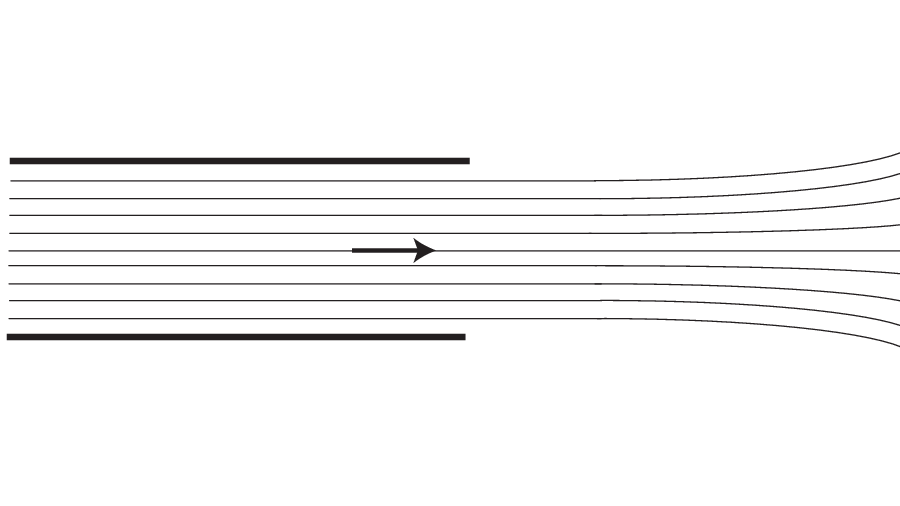}} \\
\subfigure[]{\includegraphics[width=0.28 \textwidth]{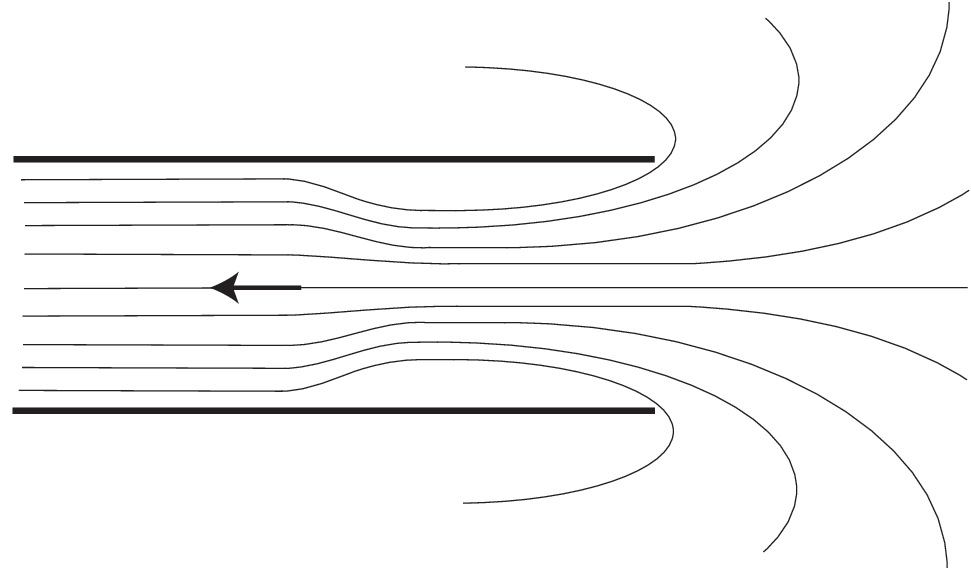}}
\caption{Steady streamlines for a real flow when (a) expelled from the mouth of a tube, (b) aspirated into the mouth of the tube.\label{fig:flows}}
\end{figure}

Mach noted that there is an asymmetry between the outflow and the inflow, since the outflow forms a narrow jet, whereas the inflow comes in from all directions, as shown in \Fig{fig:flows}.  Various authors have claimed that the omnidirectionality of the inflow is the reason why its momentum does not cancel the momentum of the outgoing jet.\cite{Baker,Gleick,Walker,Levi}  It is well-known that ``a match can be extinguished by blowing, but not by sucking,'' because the air only forms a narrow jet when blown out, and that this is related to the fluid's viscosity.\cite{Batchelor-match}

When applied to Machian propulsion, however, this line of argument can be misleading.  The asymmetry in the shape of the flows in \Fig{fig:flows} is tied to viscosity, but Machian propulsion is not.\cite{Machian}  It should be clear from our simple account of how a squid swims that Machian propulsion is simply a consequence of momentum conservation and can be explained without considering the precise shapes of the inflow and the outflow.  But, in order to understand the nature of the irreversibility manifested as Machian propulsion, we need a more careful analysis of the physics behind the flows in \Fig{fig:flows} than what is available in the standard literature.

\section{Ideal jets}
\label{sec:ideal-jets}

\begin{figure}[t]
\begin{center}
	\includegraphics[width=0.3 \textwidth]{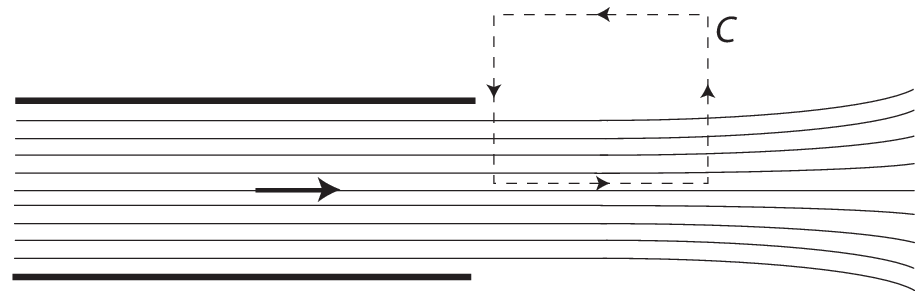}
\end{center}
\caption{The integral of $d \vv s \cdot \vv v$, along the path $C$, does not vanish for a jet flow like the one pictured in \Fig{fig:flows}(a).\label{fig:path-jet}}
\end{figure}

Kelvin's circulation theorem establishes that ideal flows (i.e., those without viscosity) that are irrotational at some initial time remain irrotational for all future times.  That is, if the flow is characterized by a velocity field $\vv v (t, x, y, z)$, then
\be
\vv \nabla \times \vv v = 0
\la{eq:irrotational}
\ee
for all $t$.\cite{dry-water}  By Stokes's theorem, this implies that
\be
\oint_C d \vv s \cdot \vv v = 0
\la{eq:pathC}
\ee
for any closed path $C$, defined within a region of space filled with a homogenous ideal fluid.

The integral on the left-hand side of \Eq{eq:pathC} is known as the ``circulation.''  It is clear from \Fig{fig:path-jet} that the circulation does not vanish everywhere for a jet surrounded by fluid approximately at rest.  One might therefore expect that it is necessary to take viscosity into account in order to understand jet formation.

In practice, the viscosity of a real fluid plays a key role in the formation of jets, as we shall review in \Sec{sec:real-jets}.  However, ideal jets {\it are} possible in principle.  If one is careful about deriving Kelvin's circulation theorem, it becomes clear that \Eq{eq:pathC} applies only if the fluid elements along $C$ initially formed some other closed path with vanishing circulation.  Since we are not allowed to draw paths that cross the solid tube in \Fig{fig:path-jet}, it is possible for an ideal jet to exit the tube and slide frictionlessly against the surrounding fluid at rest.\cite{Landau-Kelvin}  The jet would then be separated from the surrounding fluid by a surface at which the fluid's tangential velocity changes discontinuously.  Such a surface is called a ``vortex sheet.''\cite{vortex-sheet}  We shall have more to say about such idealized flows in \Sec{sec:vortex-sheets}.

\section{Real jets}
\label{sec:real-jets}

The role of viscosity in the formation of real jets may be confusing to a beginner.  Tritton writes that ``at low Reynolds numbers [i.e., when viscosity is {\it dominant}], the fluid from an orifice spreads out in all directions.  At high Reynolds numbers [i.e., when viscosity is mostly {\it negligible}] a jet, like a wake, is long and thin.''\cite{Tritton-jet}  But even at high Reynolds numbers, the viscosity cannot be ignored within a ``boundary layer'' close to the solid surface of the tube.

Within the boundary layer, the fluid's velocity drops to zero as one approaches the solid surface.\cite{Feynman-boundary}  When the fluid moves along an adverse pressure gradient (i.e., from lower to higher pressure), the streamlines in the boundary layer may break away from the solid, enclosing a region of slow, irregular flow, around which the flow changes direction.  This {\it boundary layer separation} is treated in detail in Ref.~\onlinecite{Tritton-separation}.

\begin{figure}[t]
\begin{center}
	\includegraphics[width=0.35 \textwidth]{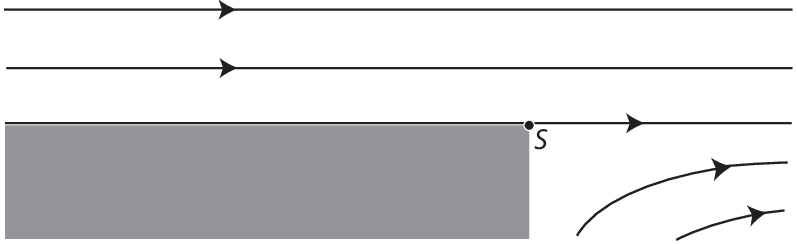}
\end{center}
\caption{Closeup of the outgoing flow in \Fig{fig:flows}(a), in the region around the lip of the tube's mouth, after it has reached a steady state.  The viscous boundary layer separates from the tube wall at $S$, where the solid has a sharp edge.  This illustration is adapted from Ref.~\onlinecite{Batchelor-separation}.\label{fig:jet-separation}}
\end{figure}

The steady flow shown in \Fig{fig:flows}(a) slows as it leaves the tube's mouth.  The pressure gradient in that region is therefore adverse.\cite{Dijkink}  Figure \ref{fig:jet-separation} illustrates the separation of the boundary layer at point $S$ on the lip of the tube's mouth.  The fluid caught in the area around $S$, where the flow changes direction, forms a thin, turbulent layer that separates the jet from the surrounding fluid.

On the other hand, the flow that enters the tube in \Fig{fig:flows}(b) always moves along a favorable pressure gradient (i.e., from high to low pressure) and the boundary layer therefore remains attached to the tube.  This explains why, for a real fluid, the outflow forms a jet, but the inflow does not.

\section{Sinks and sources}
\label{sec:sinks}

\begin{figure}[t]
\begin{center}
	\includegraphics[width=0.2 \textwidth]{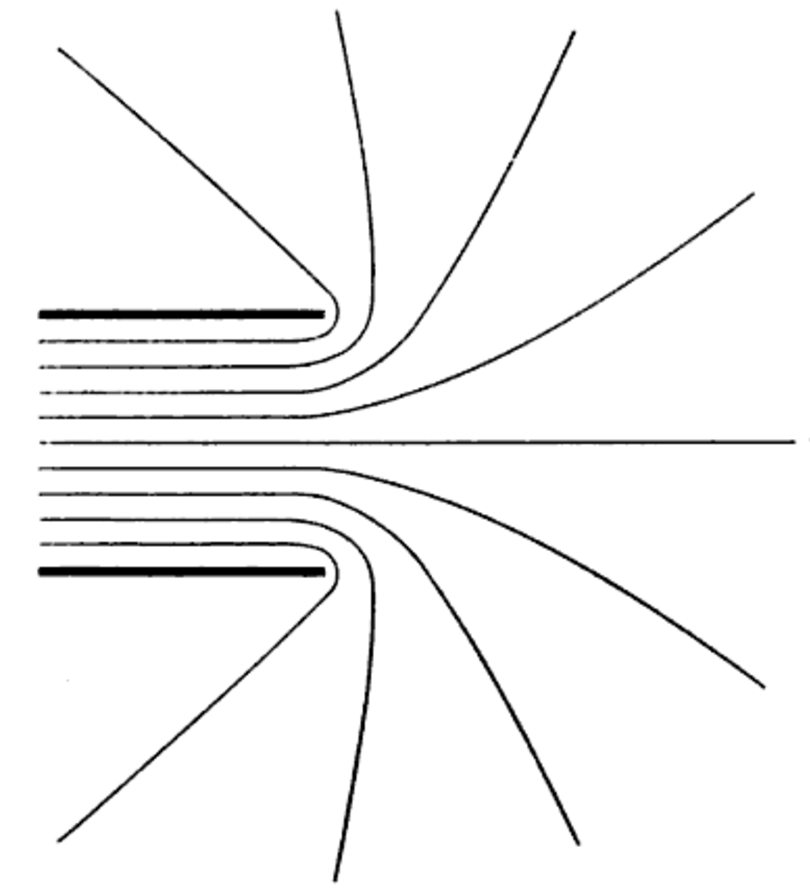}
\end{center}
\caption{Streamlines for the irrotational, incompressible, steady flow into or out of tube, within which the flow is parallel.  The spacing of adjacent streamlines corresponds to an equal flux.  The image is by Sir Horace Lamb.\cite{Lamb-Laplace}\label{fig:Laplace}}
\end{figure}

If the flow is irrotational (\Eq{eq:irrotational}), the velocity can be expressed as the gradient of a ``velocity potential'' $\varphi$:
\be
\vv v = \vv \nabla \varphi ~.
\ee
The condition of incompressibility ($\vv \nabla \cdot \vv v = 0$) is then equivalent to Laplace's equation
\be
\nabla^2 \vv \varphi = 0 ~,
\la{eq:Laplace}
\ee
whose solutions have been throughly investigated in a variety of contexts.  If the steady flow within the tube is parallel and uniform, the only possible configuration consistent with \Eq{eq:Laplace} is the one shown in \Fig{fig:Laplace}, regardless of whether the flow is entering or leaving the tube.\cite{Pippard-source}

Far from the tube, this approaches the shape for a pure sink or source of flow, with the velocity of the flow becoming radial and falling off as $1/r^2$ (in three dimensions).\cite{Batchelor-sink}  This pattern of flow is very different from the outflowing jet of \Fig{fig:flows}(a).  On the other hand, it is similar to the omnidirectional inflow in \Fig{fig:flows}(b), especially far from the tube.  We therefore refer to the flow of \Fig{fig:flows}(b) as ``sink-like.''  (In \Sec{sec:Borda} we will discuss why, unlike in \Fig{fig:Laplace}, the flow within the tube in \Fig{fig:flows}(b) is not everywhere parallel.)

As explained in \Sec{sec:real-jets}, real fluids are commonly seen to sustain sink-like but not source-like flows, because the latter are destroyed by the separation of the boundary layer at the lip of the tube's mouth.  But even in the ideal limit, a source-like flow would be difficult to set up and maintain, because it would require that the experimenter control the fluid pressure not just within the tube, but also at the boundary of the tank that encloses as the entire fluid, as we will explain in \Sec{sec:transverse}.

\section{Time reversal}
\label{sec:reversal}

Euler's equation for an ideal fluid of density $\rho$ moving under the action of a pressure field $P$,
\be
\left( \frac{\partial}{\partial t} + \vv v \cdot \vv \nabla \right) \vv v = - \frac{\vv \nabla P}{\rho}~,
\la{eq:Euler}
\ee
is invariant under the time-reversal transformation
\be
\vv v (t, x, y, z) \to - \vv v (-t, x, y, z) ~.
\la{eq:reverse}
\ee
For the most part, we will focus on {\it steady} flows, in which the velocity of the fluid moving past a given point in space remains constant (i.e., $\partial \vv v / \partial t = 0$).  For any steady flow $\vv v$ that solves Euler's equation, the reverse flow $- \vv v$ will be a solution as well, with the same pressure gradient $\vv \nabla P$.  Thus, for an ideal fluid, both the sink-like and the source-like flows imply the same pressure gradient, with higher pressure outside the tube than inside it.\cite{reverse-sprinkler}

\section{Pressure control}
\label{sec:pressure}

A mathematician thinks of \Eq{eq:Euler}, together with the condition of incompressibility, as determining both $\vv v$ and $\vv \nabla P$ (which are related to each other by Bernoulli's theorem), given a choice of initial and boundary conditions on $\vv v$.\cite{NS-math,Clay}  In practice, however, an experimenter usually establishes a flow by manipulating certain pressure differences.  This is also the case of the squid in \Fig{fig:squid}: the animal generates an outgoing jet (and therefore a propulsion), by contracting its mantle cavity and thereby increasing the pressure of the water inside it.  It then refills the cavity by expanding it and thus lowering the water pressure within it.

Note that reversing the pressure gradient in \Eq{eq:Euler}
\be
\vv \nabla P \to - \vv \nabla P
\ee
does {\it not} reverse the flow, something that at first might seem counterintuitive.  The velocity field $\vv v$ is defined as a function of time $t$ and absolute position $(x,y,z)$, but the fluid elements (to which Newton's laws apply directly) do {\it not} have fixed positions, since they move along the streamlines.  As is clearly explained in Ref.~\onlinecite{dry-water}, this is the reason for the second term in the left-hand side of \Eq{eq:Euler}, as well as the source of most of what appears confusing to a student who confronts fluid mechanics for the first time.  Reversing the force exerted by the pressure gradient on individual fluid elements changes the streamlines, rather than merely reversing their direction.

\section{Borda mouthpiece}
\label{sec:Borda}

\begin{figure}[t]
\centering
\subfigure[]{\includegraphics[width=0.25 \textwidth]{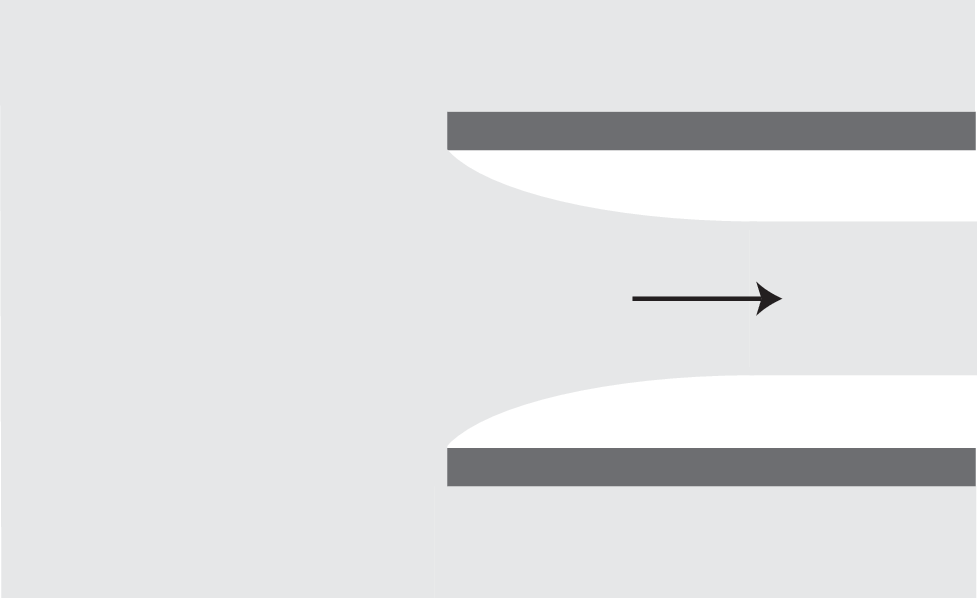}} \\
\subfigure[]{\includegraphics[width=0.25 \textwidth]{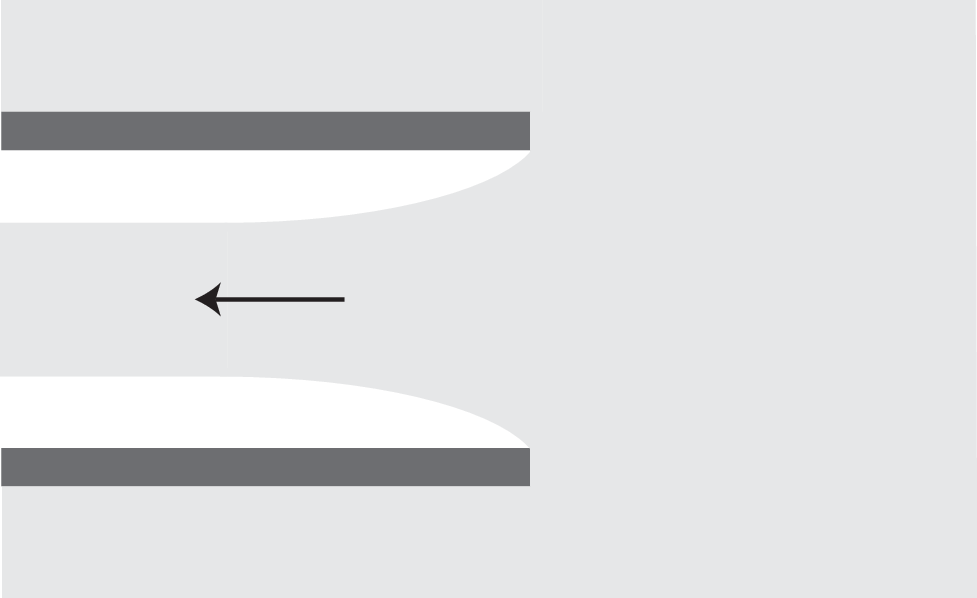}}
\caption{Ideal flow into a Borda mouthpiece forms a {\it vena contracta}, with half the tube's cross-section.  The white area within the tube may be filled with fluid at rest, against which the ideal flow (in light grey) slides frictionlessly.  The pressure gradient $\vv \nabla P$ is trivially reversed between (a) and (b), by reversing the spatial coordinates.  In both cases the pressure is lower inside the tube than in the bulk.\label{fig:Borda}}
\end{figure}

Consider the flow into a submerged tube, as shown in \Fig{fig:Borda}(a).  The fluid is accelerated by the difference between the higher pressure in the bulk and the lower pressure inside the tube.  It is easy to show, using momentum conservation and Bernoulli's theorem, that the flow within the tube forms a {\it vena contracta} (``contracted vein'') with half the tube's cross section.  The ideal flow would slide frictionlessly on the fluid at rest that surrounds the {\it vena contracta}.  For real fluids, viscous drag causes the streamlines to diverge, until the flow fills the tube and becomes parallel again, as shown in \Fig{fig:flows}(b).  The physics of this system, called a ``Borda mouthpiece,'' is reviewed in detail in Ref.~\onlinecite{Machian}.

We could trivially reverse $\vv \nabla P$ by a coordinate transformation $(x,y,z) \to (-x,-y,-z)$, but this would also reverse the position of the tube: we would merely have the fluid going out the other way, as in \Fig{fig:Borda}(b).  If we made the pressure of the fluid {\it inside the tube} higher than the pressure in the bulk, the flow pattern would have to change.  This is obvious from the fact that (by Bernoulli's theorem and the condition of incompressibility) the flow must narrow as the pressure drops.  We will describe the resulting ``ideal jet'' in \Sec{sec:vortex-sheets}.

\section{Transverse pressure gradients}
\la{sec:transverse}

\begin{figure}[t]
\begin{center}
	\includegraphics[width=0.3 \textwidth]{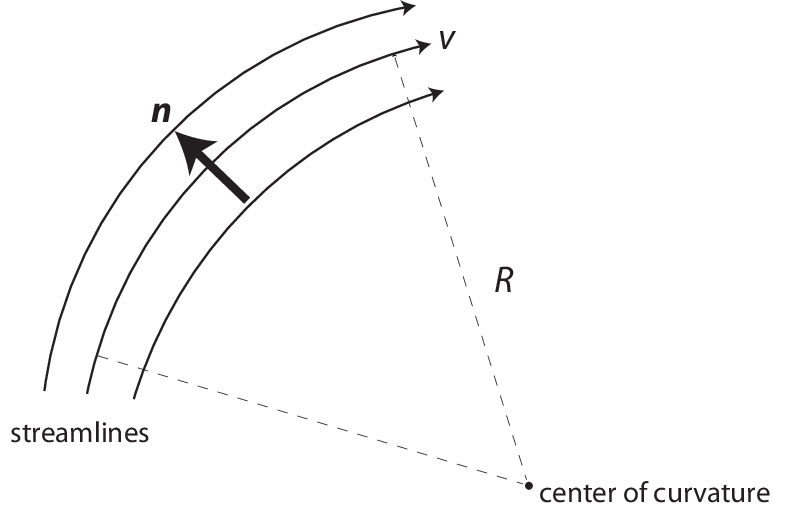}
\end{center}
\caption{When the streamlines in a flow turn with radius of curvature $R$, the pressure gradient along a unit vector $\vv n$ normal to the streamlines is given locally by \Eq{eq:steering}.  Pressure increases as one moves {\it away} from the center of curvature.\label{fig:steering}}
\end{figure}

For the streamlines of a flow to curve, there must be a pressure gradient along the direction normal to the streamlines, as illustrated in \Fig{fig:steering}.  It is the force exerted by this pressure gradient that steers the flow along the curving streamlines, overcoming the fluid's inertia.  If $\vv n$ is a unit vector normal to the streamlines of the flow at a given point, we have that
\be
\vv n \cdot \vv \nabla P = \frac{\rho v^2}{R}~,
\la{eq:steering}
\ee
where $R$ is the radius of curvature of the streamlines.\cite{Babinsky}  For the flow in \Fig{fig:Laplace}, in which $R$ is very small near the lip of the tube's mouth, \Eq{eq:steering} implies that the pressure of the fluid near the edge of the tube's mouth has to be much smaller than the pressure of the fluid that is nearly at rest along the boundary of the tank.

\begin{figure}[t]
\begin{center}
	\includegraphics[width=0.3 \textwidth]{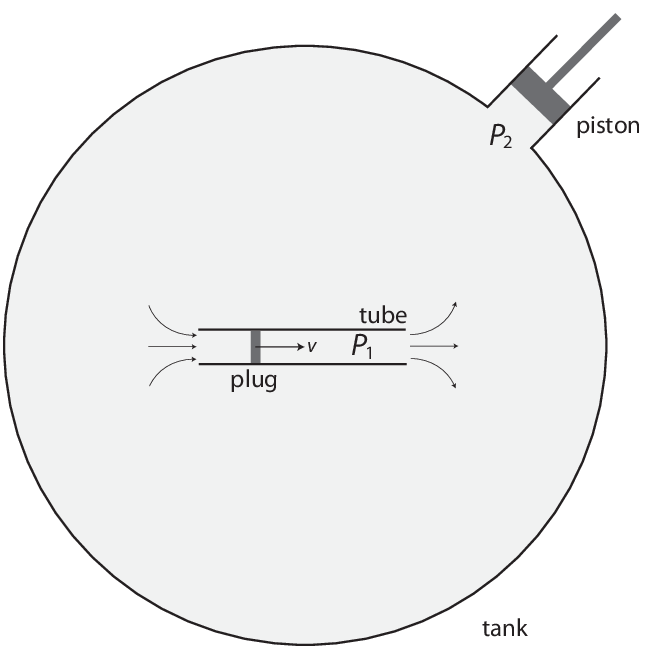}
\end{center}
\caption{A long and narrow tube, placed inside a large tank filled with fluid, is open at both ends and contains a movable plug.  The pressure at the tank's boundary can be controlled by a piston.  This diagram is adapted from an illustration by Sir Horace Lamb.\cite{Lamb-sink-source}\label{fig:Lamb-sink-source}}
\end{figure}

Consider a long and narrow tube, open at both ends, placed inside of a tank filled with an ideal fluid.  Let the tube contain a movable plug.  If this plug is moved along the tube with a constant speed $v$, then the only possible steady, irrotational flow that can result is the one shown in \Fig{fig:Lamb-sink-source}, with a sink-like flow into one end of a tube and a source-like flow out of the other end.  The pressure of the parallel flow within the tube is labelled $P_1$.  As the flow leaves the tube's mouth and spreads, its speed decreases and its pressure increases.  By Bernoulli's theorem, the fluid at rest near the boundary of the tank will have pressure
\be
P_2 = P_1 + \frac{1}{2} \rho v^2 ~.
\la{eq:Bernoulli}
\ee

Lamb analyzed this configuration in Ref.~\onlinecite{Lamb-sink-source}, from where we have taken \Fig{fig:Lamb-sink-source}.  As he notes, maintaining this pattern of flow as the plug is moved down the tube would require that the experimenter maintain the higher pressure $P_2$ at the boundary of the tank, which could be done by pushing on a piston like the one shown in \Fig{fig:Lamb-sink-source}.  But if the experimenter lets the piston recede, the streamlines of the fluid that leaves the tube would not be able to spread outwards, because there would not be a sufficient pressure gradient to steer the streamlines around the tube's mouth.  In that case, an annular cavity might form at both ends of the tube.  A cavity (or bubble) is a region that contains no fluid and therefore effectively exerts a negative pressure.\cite{Batchelor-cavity}

\section{Pump-driven flows}
\la{sec:pumps}

So far, nothing in our analysis has challenged the symmetry between ideal inflows and outflows implied by the Euler equation.  In the process represented by \Fig{fig:Lamb-sink-source}, a given fluid element moves from a region of higher pressure $P_2$ to a region of lower pressure $P_1$ and finally back to $P_2$.  The direction of this motion can be reversed by reversing the plug's motion.  But Lamb's analysis reveals that the shape of the flow around the tube's openings depends on the pressure at the tank's boundary.

\begin{figure}[t]
\begin{center}
	\includegraphics[width=0.3 \textwidth]{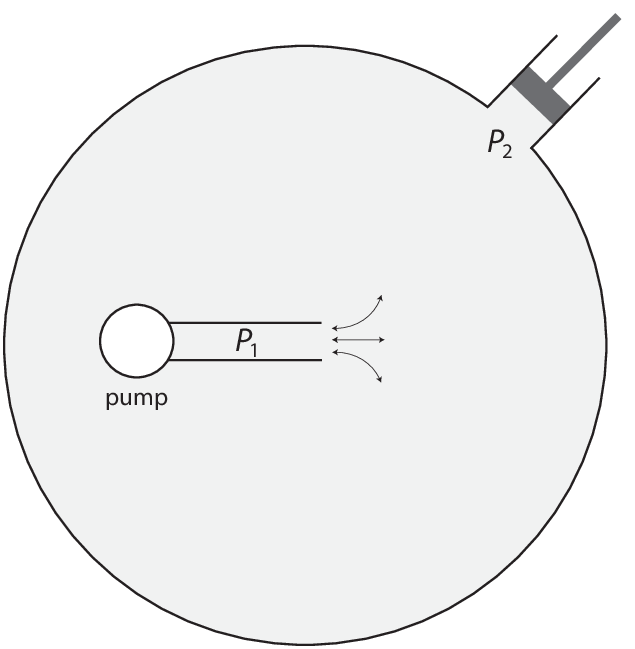}
\end{center}
\caption{Instead of the configuration in \Fig{fig:Lamb-sink-source}, we now consider attaching one end of the tube to a pump.  It is easy to establish a sink-like flow by lowering the pump's pressure, but maintaining a source-like flow would require control of the pressure $P_2$ at the tank's boundary.\label{fig:pump}}
\end{figure}

Suppose that, instead of moving a plug, the experimenter controlled a pump attached to one end of the tube inside the tank, as shown in \Fig{fig:pump}.  Fluid can be expelled by increasing the pressure in the pump relative to the fluid at rest in the tank.  Conversely, fluid may be sucked in by lowering the pump pressure.  This is the case in all of the systems that exhibit Machian propulsion, such as the squid in \Fig{fig:squid}.

It is easy to establish a sink-like flow with such a device.  If the pump keeps the flow within the tube at a pressure $P_1$ lower than the tank pressure $P_2$, fluid will rush omnidirectionally from the tank into the tube, forming a {\it vena contracta} (as discussed in \Sec{sec:Borda}).  But producing a source-like flow is more difficult.  With an arrangement like the one in \Fig{fig:pump}, fluid can only be pushed down the tube by first rising the pump pressure relative to the tube's $P_1$.  Once the flow in the tube has been established, the experimenter would have to act on the piston in order to maintain a higher pressure $P_2$ near the tank's inner boundary, in accordance with \Eq{eq:Bernoulli}.  Otherwise, the streamlines leaving the tube's mouth would not curve outwards.  In \Sec{sec:vortex-sheets} we will describe what would happen to the ideal flow in that case.

Evidently, an experimenter who controls only the pump in \Fig{fig:pump} will be unable to reverse the flow in the tank.  As we explained in \Sec{sec:machian}, the inflow and outflow phases of Machian propulsion, in which the sign of the pressure difference is flipped, are not time-reversed images of each other, even in the limit of zero dissipation.

In the case of the source-like flow, Bernoulli's theorem implies that the pressure $P_2$ at the tank's inner boundary must equal the pressure within the pump (where the fluid is initially at rest).  Thus, by raising $P_2$ in order to maintain the source-like flow, the experimenter would also equalize the pressures on opposite sides of the pump's wall, causing the net force on that wall to vanish.  This is compatible with the fact that, unlike a jet, the source flow transfers no net momentum to the tank walls.  Thus, by conservation of momentum, no net force would act on the solid device submerged inside the tank.\cite{transfer}  The squid would come to a standstill!

\section{Vortex sheets and cavities}
\la{sec:vortex-sheets}

As we discussed in \Sec{sec:ideal-jets}, an ideal flow rushing out of a solid tube could form a jet, separated from the surrounding fluid by a ``vortex sheet,'' at which the tangential velocity changes discontinuously.  Though useful in certain contexts,\cite{Anderson-vortex} such vortex sheet solutions to Euler's equation are unstable: small transverse perturbations grow exponentially until the nonlinearities associated with turbulence become important and the flow cannot be treated as ideal, even approximately.\cite{Landau-Kelvin,KH-instability}  It is instructive, nonetheless, to consider such idealized flows in order to explain why Machian propulsion, which appears as a kind of irreversibility, is nonetheless compatible with the theory of non-dissipative flows.

\begin{figure}[t]
\begin{center}
	\includegraphics[width=0.3 \textwidth]{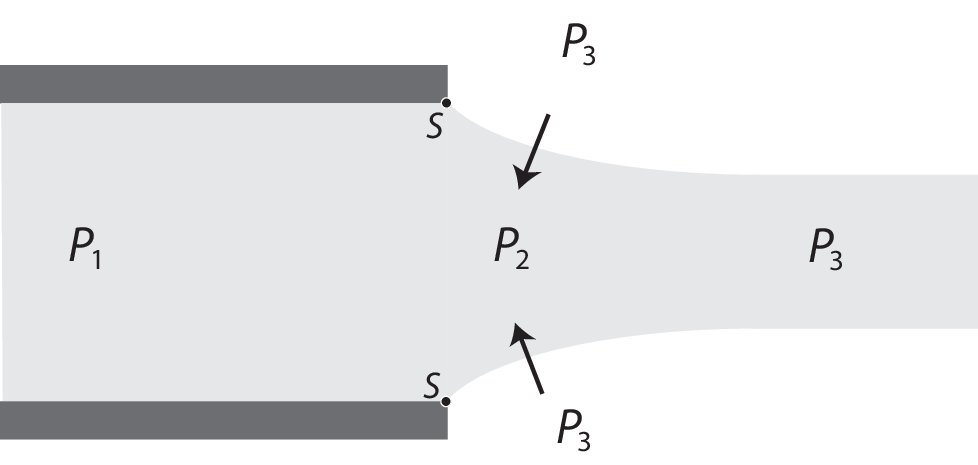}
\end{center}
\caption{An ideal, incompressible fluid at high pressure $P_1$ exits through a tube and into a region where the fluid is at rest at a {\it lower} pressure $P_3$.  Only the moving fluid is shaded.  Upon leaving the tube, the flow has an intermediate pressure $P_2$.  Since $P_2 > P_3$, there is a transverse pressure gradient in the direction of the arrows, and the streamlines therefore curve inwards in that region.  The sharp turning of the streamlines at the points marked $S$ requires the formation of a cavity around the tube's mouth.\label{fig:ideal-jet}}
\end{figure}

If the pressure within the tube is greater than the pressure outside, Bernoulli's theorem and the condition of incompressibility imply that the outgoing jet must be narrower than the tube (unlike the flow shown in \Fig{fig:path-jet}).  The pressure $P_2$ of the jet leaving the tube is higher than the ambient $P_3$ of the surrounding fluid at rest.  As shown in \Fig{fig:ideal-jet}, the resulting transverse pressure gradient from $P_2$ to $P_3$ makes the streamlines converge as they leave the tube, until the pressure of the moving fluid drops to $P_3$ and the flow becomes parallel.

Note that Bernoulli's theorem cannot be applied to points in the fluid that are separated by the vortex sheet.  The reason is that that theorem expresses energy conservation, but the total energy of fluid elements in the outflow is greater than the energy of similar elements in the ambient fluid.  Thus, the same pressure $P_3$ can apply to the parallel jet flow and to the surrounding fluid at rest.\cite{Babinsky-Bernoulli}

Note also that the outflowing streamline must turn sharply inward as it passes near the point $S$ in \Fig{fig:ideal-jet}, at the lip of the tube's mouth.  This requires a very large transverse pressure gradient there.  The only way in which the flow itself can create such a gradient is by forming a cavity near $S$, where the pressure becomes negative.  Absent such a cavity, we must have $P_1 = P_2 = P_3$, so that the streamlines remain parallel as they leave the tube.  In such a case, the flow would have to be accelerated {\it before} entering the tube, by a higher pressure within the pump. 

Even a small amount of viscosity completely alters this picture, producing instead a jet like that of \Fig{fig:jet-separation}, in which the cavity at $S$ is replaced by separation of the boundary layer, while the vortex sheet is replaced by a thin layer of turbulent flow.  The theoretical details of this (which Lamb describes as ``not easy to make out''\cite{Lamb-sink-source}) need not concern us here.  Our objective has been only to establish that the irreversibility of Machian propulsion is a simple consequence of momentum conservation, not of dissipation.  A realistic solution to the more difficult problem of determining the shapes of the resulting flows must invoke viscosity, as we saw already in \Sec{sec:real-jets}.

\section{Conclusions}
\label{sec:conclusions}

The Navier-Stokes equation for real fluids,
\be
\left( \frac{\partial}{\partial t} + \vv v \cdot \vv \nabla - \nu \nabla^2 \right) \vv v = - \frac{\vv \nabla P}{\rho}
\la{eq:Navier-Stokes}
\ee
has no time-reversal symmetry.  This is understood to result from the fact that a kinematic viscosity $\nu > 0$ implies that the macroscopic energy of the flow may be irreversibly dissipated into the random thermal motion of the fluid's microscopic components.  But even an ideal flow with $\nu = 0$ can exhibit a certain kind of irreversibility.

An experimenter who controls only the pressure of a pump connected to a solid tube will see jets form when the fluid is expelled from the tube, never when the fluid is aspirated.  Conversely, the experimenter will be able to set up and maintain a sink-like flow into the tube, but not a source-like flow out of it: the pattern in \Fig{fig:Laplace} is plausible only if the streamlines point into the tube.  The reason for this is that the source-like flow would require control of the pressure of the fluid not just at the pump attached to the tube, but also at the distant boundary of the tank that receives the fluid, where the pressure must be kept higher than that inside the tube.

Alternatively, we may note that a jet ---either a real one as in \Fig{fig:flows}(a) or an ideal one as in \Fig{fig:ideal-jet}--- carries momentum away to infinity (or to the boundary of the surrounding tank).  Reversing this motion therefore requires that momentum flow {\it in} from infinity (or from the tank walls).  Setting this up would obviously demand a control of the fluid that an experimenter can not achieve with only a pump attached to the submerged tube.

This irreversibility is similar in spirit to that of statistical mechanics, which also results from the experimenter's limited control of initial and boundary conditions.  What is remarkable is that it can be seen even in an ideal flow, without invoking any concept of heat, temperature, or entropy.  A video of an ideal fluid moving past the mouth of a tube would, if played backwards, appear suspect, even though the second law of thermodynamics would play no part in the flow.

The Machian propulsion described in \Sec{sec:machian} can be entirely understood in terms of momentum conservation and is {\it most} pronounced in the limit of zero viscosity.  Even without dissipation, the outflow and inflow phases are not time-reversed images of each other: it is only the pressure difference between the inside and the outside of the solid device that changes sign.  Machian propulsion is therefore an instance of nonthermal irreversibility.

\begin{acknowledgements}

The author thanks Paul O'Gorman for help understanding the role of boundary layer separation in the formation of real jets, as well as the physical significance of the time-reversal symmetry of Euler's equation.  Vanessa L\'opez pointed him to Ref.~\onlinecite{Landau-Kelvin}, while Giancarlo Reali provided general feedback on the manuscript.  The author also benefitted from discussions with Carl Mungan and Take Okui.  An anonymous referee's comments contributed significantly to clarifying this presentation.  Tim Lowson, of Ready-Ed Publications, kindly gave permission to use \Fig{fig:squid}.  This work was supported in part by the U.S.~Department of Energy under contract DE-FG02-97IR41022.

\end{acknowledgements}


\bibliographystyle{aipprocl}   

\begin{thebibliography}{99}
\bibitem{transfer}
This is consistent with the simple analysis of Ref.~\onlinecite{Reverse}, where the force on the submerged solid was decomposed into a contribution from ``pressure difference'' and another from ``momentum transfer.''  Note, however, that when a flow impinges on a solid wall and changes direction, \Eq{eq:steering} implies that there will be an associated pressure gradient.  This change of pressure is what actually accounts for momentum transfer (i.e., for the force that the flow exerts on the solid on which it impinges).
\bibitem{Babinsky-Bernoulli}
This point is very lucidly made in Ref.~\onlinecite{Babinsky}, where the author notes that it renders unsound the popular explanation of aerodynamic lift in terms of Bernoulli's theorem.
\end{thebibliography}

\end{document}